\documentclass[aps,pra,superscriptaddress,twocolumn,longbibliography]{revtex4-1}
\usepackage{mathrsfs}
\usepackage{amsfonts}
\usepackage{amssymb}
\usepackage{amsmath}
\usepackage{graphicx}

\graphicspath{{figures/}{./}} 
\usepackage{sidecap}
\usepackage{color}
\usepackage{xcolor}
\usepackage[colorlinks=true, urlcolor=blue, citecolor=blue,linkcolor=blue,citebordercolor={1 0 0},linkbordercolor={0 0 1}]{hyperref}
\usepackage{subeqnarray}
\usepackage{verbatim}
\usepackage{euscript} 

\usepackage{hyperref}

\usepackage{bibunits}

\usepackage{array}
\usepackage{multirow}

\usepackage{braket}

\renewcommand{\eqref}[1]{Eq.~(\ref{#1})} 
\newcommand{\figref}[1]{Fig.~\ref{#1}} 

\newcommand{\harvard}{Department of Physics, Harvard University, Cambridge, MA 02138, USA}
\newcommand{\rle}{Research Laboratory of Electronics, Massachusetts Institute of Technology, Cambridge, MA 02139, USA}
\newcommand{\mitphysics}{Department of Physics, Massachusetts Institute of Technology, Cambridge, MA 02139, USA}
\newcommand{\miteecs}{Department of Electrical Engineering and Computer Science, Massachusetts Institute of Technology, Cambridge, MA 02139, USA}

\newcommand{\mitll}{Lincoln Laboratory, Massachusetts Institute of Technology, Lexington, MA 02421-6426, USA}

\newcommand{\ut}{Departments  of  Chemistry and Computer science,  University  of  Toronto,  Toronto,  Ontario  M5G 1Z8,  Canada}
\newcommand{\vectorinst}{Vector  Institute  for  Artificial  Intelligence,  Toronto,  Ontario  M5S  1M1,  Canada}
\newcommand{\cifar}{Canadian  Institute  for  Advanced  Research,  Toronto,  Ontario  M5G  1Z8,  Canada}

\begin{document}
\title{Demonstration of tunable three-body interactions between superconducting qubits}

\author{Tim Menke}
\email{timmenke@mit.edu}
\thanks{These authors contributed equally to this work.}
\affiliation{\rle}
\affiliation{\mitphysics}
\affiliation{\harvard}

\author{William P. Banner}
\thanks{These authors contributed equally to this work.}
\affiliation{\miteecs}

\author{Thomas R. Bergamaschi}

\author{Agustin Di Paolo}
\affiliation{\rle}

\author{Antti Veps\"al\"ainen}
\affiliation{\rle}

\author{Steven J. Weber}
\affiliation{\mitll}

\author{Roni Winik}
\affiliation{\rle}

\author{Alexander Melville}
\affiliation{\mitll}

\author{Bethany M. Niedzielski}
\affiliation{\mitll}

\author{Danna Rosenberg}
\affiliation{\mitll}

\author{Kyle Serniak}
\affiliation{\mitll}

\author{Mollie E. Schwartz}
\affiliation{\mitll}

\author{Jonilyn L. Yoder}
\affiliation{\mitll}

\author{Al\'an Aspuru-Guzik}
\affiliation{\ut}
\affiliation{\vectorinst}
\affiliation{\cifar}

\author{Simon Gustavsson}
\affiliation{\rle}

\author{Jeffrey A. Grover}
\affiliation{\rle}

\author{Cyrus F. Hirjibehedin}
\affiliation{\mitll}

\author{Andrew J. Kerman}
\affiliation{\mitll}

\author{William D. Oliver}
\email{william.oliver@mit.edu}
\affiliation{\rle}
\affiliation{\mitphysics}
\affiliation{\miteecs}
\affiliation{\mitll}

\date{\today}

\begin{abstract}
Nonpairwise multi-qubit interactions present a useful resource for quantum information processors.
Their implementation would facilitate more efficient quantum simulations of molecules and combinatorial optimization problems, and they could simplify error suppression and error correction schemes.
Here we present a superconducting circuit architecture in which a coupling module mediates 2-local and 3-local interactions between three flux qubits by design.
The system Hamiltonian is estimated via multi-qubit pulse sequences that implement Ramsey-type interferometry between all neighboring excitation manifolds in the system.
The 3-local interaction is coherently tunable over several MHz via the coupler flux biases and can be turned off, which is important for applications in quantum annealing, analog quantum simulation, and gate-model quantum computation.
\end{abstract}

\maketitle


A key challenge in the development of quantum computers is the implementation of resource-efficient and precisely tunable interactions between qubits \cite{kjaergaard2020superconducting}.
To date, most of the interactions that have been implemented in quantum systems are pairwise in nature.
While pairwise interactions, which are referred to as 2-local, are sufficient to generate entanglement across a many-qubit system \cite{barends2014superconducting, bernien2017probing, landsman2019verified}, there are many cases, particularly when using limited-depth circuits, in which such interactions are insufficient or inconvenient:
multi-qubit interactions are a prerequisite for analog quantum simulations of chemistry Hamiltonians and certain condensed matter physics models \cite{babbush2014adiabatic, Majumdar2012} as well as for quantum annealing for combinatorial optimization \cite{Marto_k_2004, Santoro_2006, hauke2020perspectives}.
They play a key role in error suppression schemes \cite{Bacon2006} and in parity checks for error correction algorithms \cite{fowler2012surface}.

Experimental demonstrations of multi-qubit interactions are scarce:
a 4-local ring exchange has been observed in a cold-atom system when suppressing lower-order interactions \cite{dai2017four}, and a small, chiral 3-local interaction has been engineered between dynamically driven superconducting qubits \cite{liu2020synthesizing}.
Thus far, the interactions have been slow and not suitable for use in a scalable quantum information processing architecture.
In addition, few metrological methods exist to extract all interactions of a nonpairwise coupled system precisely \cite{bergamaschi2021distinguishing}.
However, significant interest in multi-qubit coupling mechanisms persists, as evidenced by a number of proposals for tunable multi-qubit couplers for quantum processors \cite{Mezzacapo_2014, Hafezi_2014, chancellor2017circuit, Schondorf_2019, melanson2019tunable, menke2019automated}.

In this work, we demonstrate tunable 3-local interactions between superconducting flux qubits.
The interactions are mediated by a coupler circuit, which enables static coupling without the need for dynamic driving.
A multi-qubit Hamiltonian estimation technique is implemented to determine the system parameters:
the coherence of the qubits, which is drastically improved over typical annealing-type qubits, is exploited to implement multi-qubit Ramsey sequences for precise metrology of the system eigenenergies.
This technique distinguishes the 3-local coupling from each individual 2-local interaction between the qubits.
We find that the 3-local coupling strength can be tuned from an essentially off bias point to a maximal strength of $-6.5\,$MHz, which is comparable to typical interaction rates in certain state-of-the-art digital processors \cite{sung2021realization}.
Numerical simulations of the full circuit Hamiltonian elucidate the coupling mechanism and show that it arises from interactions between the coupler excited state and higher excited states of the qubit system.
The coupling is also tunable by about 3\,MHz along a flux insensitive path in the coupler dispersion, preserving maximum qubit coherence.
Therefore, our work presents both a demonstration of an elusive coupling mechanism and a solution for more resourceful interactions in quantum processors.

\begin{figure}[htb]
\centering
\includegraphics[scale=1.0]{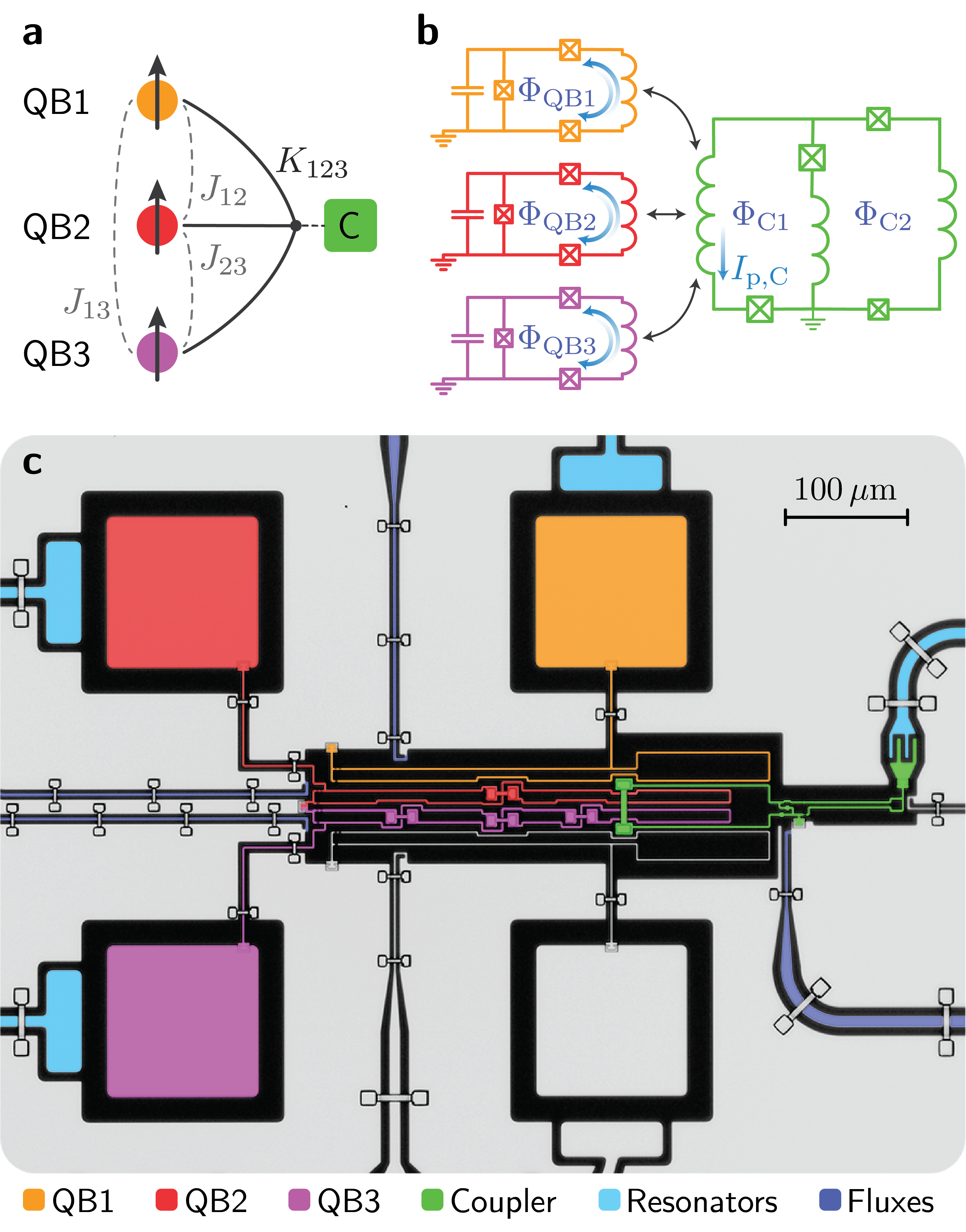}
\caption{
Coupled multi-qubit system. (a) The qubits are coupled to a common coupler C, which mediates a 3-qubit interaction. Spurious terms of lower locality are included in the model (dashed lines). (b) Superconducting circuit that implements the desired system. It is constructed from flux qubits that couple inductively and capacitively to the coupler circuit. External fluxes $\Phi$ determine the local spin fields and coupler properties. (c) Micrograph of the on-chip realization of the circuit.
}
\label{fig:intro}
\end{figure}


We consider a system of three qubits that are pairwise coupled both among themselves and to a coupler element (see \figref{fig:intro}(a)).
The qubits are modeled as spins that are aligned or anti-aligned with the $z$-axis, with the ground (excited) state given by $\ket{0}$ $\left(\ket{1}\right)$.
The coupler element induces a nonpairwise 3-local interaction as well as spurious 2-local interactions, which add to the existing capacitive and inductive 2-local interactions.
As the coupler has a larger frequency gap than the qubits and is not excited during operations, we can model the system as an effective 3-qubit system with the following eigenbasis Hamiltonian:
\begin{equation}
\label{eq:spin_hamiltonian}
    H/\hbar = -\sum_{i=1}^{3} \frac{\omega_i}{2} \hat{Z}_i + \sum_{\substack{i,j=1 \\ i<j}}^{3} J_{ij} \hat{Z}_i \hat{Z}_j + K_{123} \hat{Z}_1 \hat{Z}_2 \hat{Z}_3,
\end{equation}
where $\omega_i$ denotes the single-qubit frequencies, $J_{ij}$ the 2-local and $K_{123}$ the 3-local coupling strengths.
The Pauli $\hat{Z}$ matrix for qubit $i$ is given by $\hat{Z}_i$.

The system is implemented as a superconducting circuit that consists of three flux qubits \cite{orlando1999superconducting, you2007low, yan2016flux} and a flux tunable coupler \cite{menke2019automated, menke2022inprep} (see \figref{fig:intro}(b)).
The flux qubit eigenstates are superpositions of clockwise and counterclockwise circulating currents in the qubit loop.
Throughout this work, the qubits are operated at the flux insensitive point, which occurs when the external flux threading the loop is $\Phi_{\text{QB}i} = 0.5\,\Phi_0$.
At this flux bias, the three qubit frequencies $\Delta_{\text{QB}i}$ are in the range of 2.5--5.5\,GHz.
Before diagonalizing the system to obtain the form of \eqref{eq:spin_hamiltonian}, each flux qubit is described by the following Hamiltonian in the persistent-current basis:
\begin{equation}
    H_{\text{QB}i} = \varepsilon\left( \Phi_{\text{QB}i} \right) \hat{z}_i + \Delta_{\text{QB}i} \, \hat{x}_i,
\end{equation}
where $\hat{z}_i$ and $\hat{x}_i$ are the Pauli matrices specified in the persistent-current basis and the flux insensitive point is parameterized to $\varepsilon = 0$.
The pairwise coupling to other qubits and the coupler includes both inductive and capacitive interactions.

The coupler circuit was first proposed in Refs.~\cite{kerman2018design, menke2019automated} and shown experimentally to exhibit a nonlinear coupling potential versus flux \cite{menke2022inprep}.
It is therefore expected to mediate nonpairwise interactions when inductively coupled to a set of flux qubits.
At the same time, the excited state of the coupler can push down the qubit energy levels and induce effective interactions in the qubit subspace when biased around its minimum frequency gap of about 9\,GHz.
In order to predict the system couplings, we numerically simulate the circuit Hamiltonian in a mixed representation of the charge and harmonic oscillator bases, using a hierarchical diagonalization strategy to solve the qubit and coupler Hamiltonians separately before adding the interactions \cite{kerman2020efficient}.

The circuit was fabricated with a high-quality aluminum-based process patterned on a silicon substrate, embedded in readout and control infrastructure as shown in \figref{fig:intro}(c) \cite{yan2016flux}.
The qubit loops are elongated to encompass surface area within or in proximity to the left coupler loop, thereby coupling the qubits inductively and capacitively to the coupler.
In two of the qubit loops, bowtie-shaped crossovers are used to route wires over one another and connect ground planes and twist the loops \cite{rosenberg2020solid}. These twists reduce the inductive coupling between the qubits.
Individual readout resonators are capacitively coupled to the shunt capacitor of each qubit, with a state-dependent resonator shift $\chi_i$ between 2--20\,MHz at the flux insensitive point and coupling rate $\kappa_i = \text{1--2\,MHz}$ to a shared feedline, which enables fast readout with a 360\,ns integration time.
Qubit operations are implemented by resonantly driving the qubit through the resonator.
Local flux lines permit control of the individual loop fluxes.
Flux crosstalk to non-primary loops is suppressed to a mean of 0.5\% and maximum of 3.4\% between any antenna-loop pair via an iterative calibration procedure based on Refs.~\cite{dai2021calibration, menke2022inprep} (see Supplementary Information).
The on-chip circuit includes a fourth flux qubit that is not needed and so was far detuned from the other transitions.

\begin{figure}[htb]
\centering
\includegraphics[scale=1.0]{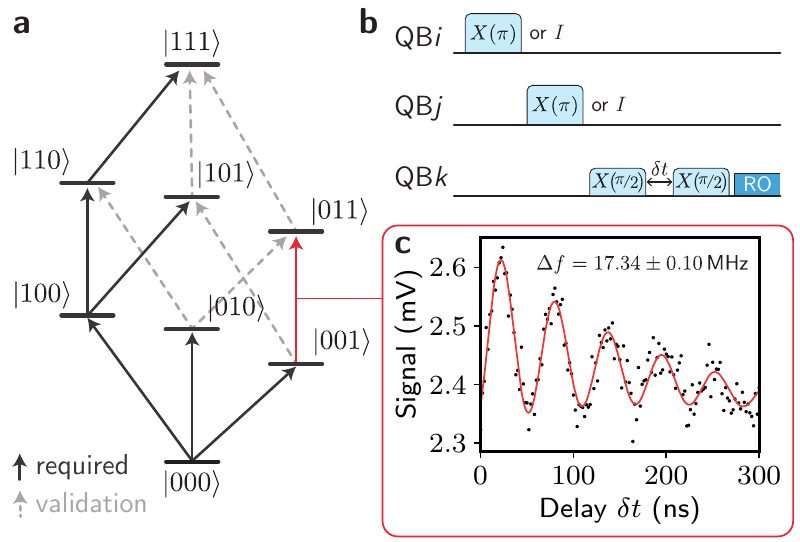}
\caption{
Hamiltonian estimation method.
(a) Energy level diagram of the 3-qubit system. The allowed single-photon transitions are indicated with arrows. The 3-qubit Hamiltonian model is fully determined by a subset of transitions (solid black). The other transitions are used to validate the model.
(b) Multi-qubit pulse sequence that implements Ramsey interferometry between arbitrary single-photon transitions.
(c) Example Ramsey interferometry data for the $\ket{001} \rightarrow \ket{011}$ transition. The fit is used to determine the transition frequency precisely, which is determined by the sum of the drive frequency and the extracted detuning $\Delta f$.
}
\label{fig:detection_method}
\end{figure}

In order to estimate the Hamiltonian of the 3-qubit system and extract its interactions, we measure the eigenenergies of the system up to a total of three excitations, one per qubit.
The transitions between eigenstates in adjacent excitation manifolds are shown in \figref{fig:detection_method}(a).
Each transition frequency is a linear combination of the Hamiltonian parameters in the eigenbasis of the full system.
If we are able to identify and precisely measure a set of transitions that include each eigenstate at least once, the Hamiltonian parameters in \eqref{eq:spin_hamiltonian} can be determined by inverting a linear system of equations.
We use a Ramsey interferometry experiment to determine each transition frequency \cite{ramsey1950molecular}, for example for the transition $\ket{001} \rightarrow \ket{011}$ (see \figref{fig:detection_method}(b,c)).
It is realized by applying successive $\pi$-pulses such that the system is driven from the ground state to the lower state of the transition of interest, which in the example case requires the sequence
\begin{equation}
\ket{000} \overset{X_3(\pi)}{\longrightarrow} \ket{001}.
\end{equation}
The Ramsey experiment then proceeds with $\pi/2$-pulses slightly detuned from the desired transition, separated by a time delay $\delta t$ and followed by a readout pulse.
The precise transition frequency is manifest in the Ramsey fringe oscillation frequency, which is determined from a fit to the data of a sinusoid that includes the $T_1$ decay of the other qubits (see \figref{fig:detection_method}(c)).
The error is estimated as the one-sigma confidence interval for the fit parameter \cite{newville2016lmfit}.

\begin{figure}[htb]
\centering
\includegraphics[scale=1]{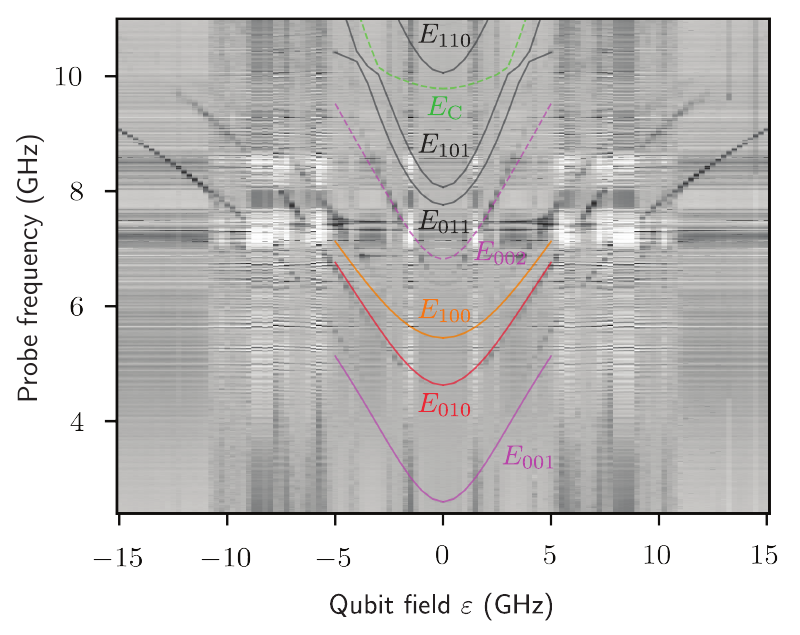}
\caption{
Spectroscopy of the system at maximum coupling around the flux insensitive point $\varepsilon = 0$ of the qubits.
The excited states of the qubits as well as some higher-lying states are visible.
Eigenenergies obtained from a full circuit Hamiltonian model are overlaid.
The data are acquired by strongly driving the circuit through a shared feedline and reading out the QB3 resonator.
}
\label{fig:spectroscopy}
\end{figure}

It is essential for the Hamiltonian estimation method that the computational eigenstates are correctly identified, which is generally a challenge for strongly coupled systems.
We first perform qubit spectroscopy to determine the lowest three transitions in the system, which correspond to single-qubit excitations.
In \figref{fig:spectroscopy}, we show a qubit spectroscopy data set that is obtained by sweeping the frequency of a continuous-wave tone and monitoring the resonator coupled to QB3.
Readout crosstalk between each qubit and the three resonators makes visible the excited states of all three qubits as well as some higher excited states.
Having identified the lowest transitions, we proceed to find successively higher transitions in the computational subspace by applying $\pi$-pulses to lower transitions and performing Rabi spectroscopy around the bare frequency of the respective qubit.
To exclude the mislabeling of undesired multi-photon transitions, we double the drive amplitude and check that the Rabi oscillation frequency also doubles.
In addition, after all transitions are identified and measured via the multi-qubit Ramsey protocol, we use a minimal subset of transitions to predict the remaining ones. 
By finding that the predictions match the measurements to within the error estimates, we have verified that the identified transitions connect a closed set of computational states.
We refer the reader to the Supplementary Information for additional details about the eigenstate verification procedure.
The set of computational states is then used to fit a full circuit Hamiltonian model to the spectroscopy data, which is overlaid with that data in \figref{fig:spectroscopy}.
The model is valid around the flux insensitive point of the qubits, which corresponds to $\varepsilon = 0$.
In addition, it approximately captures the second excited state energy $E_{002}$ of QB3 as well as the coupler excited state energy $E_\text{coup}$, which are faintly visible in the spectrum.

\begin{figure}[htb!]
\centering
\includegraphics[scale=1]{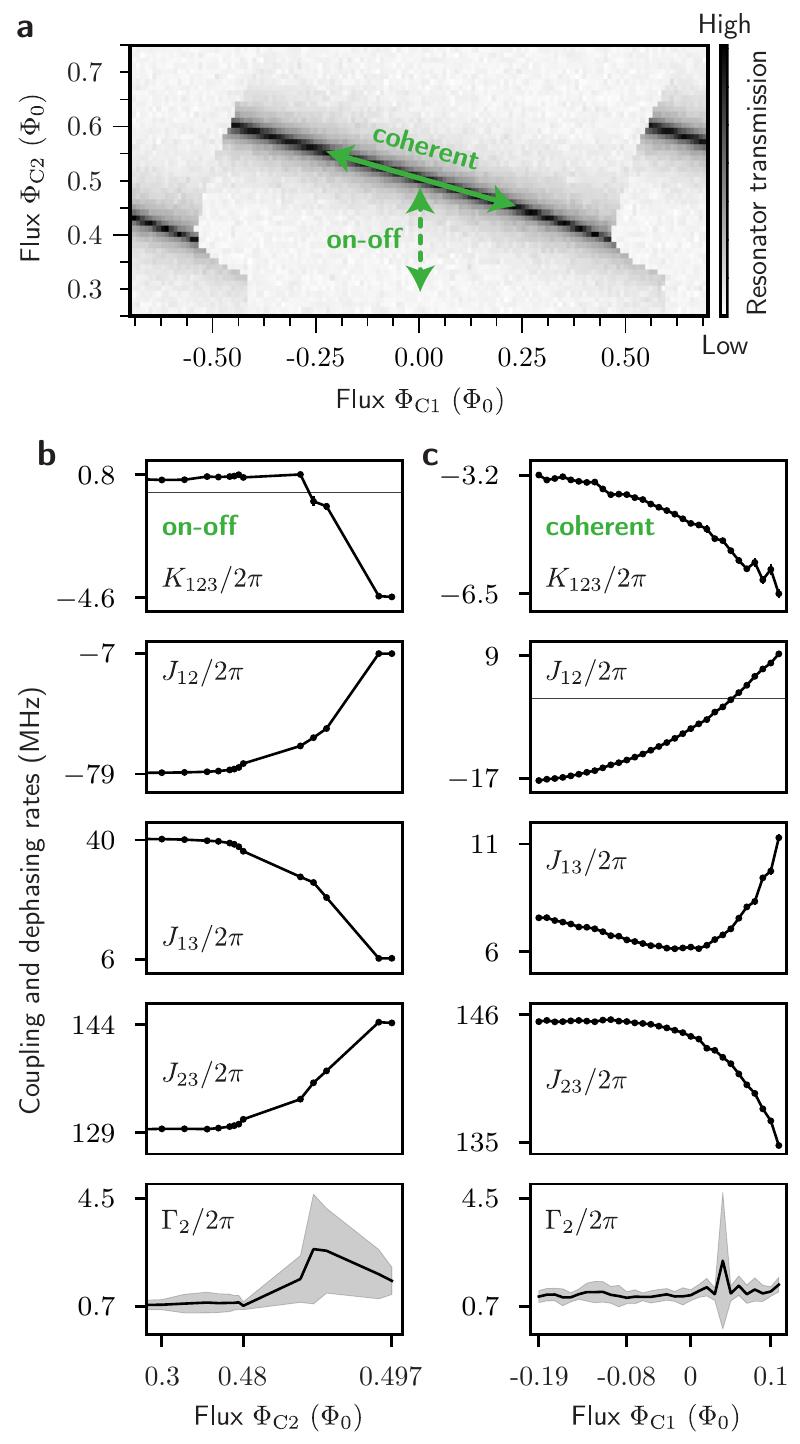}
\caption{
Tuning of the qubit interactions with coupler flux bias.
(a) Transmission spectrum of the coupler resonator versus the coupler fluxes.
Transmission is measured at a fixed frequency below the bare resonator frequency.
(b) Tunability of the interactions is largest along the on-off direction.
Shown are the 3-local interaction $K_{123}$, the 2-local interactions $J_{ij}$, and the mean decoherence rates $\Gamma_2$.
Error bars for the couplings are smaller than the markers unless visible, and the shaded area for $\Gamma_2$ indicates the standard deviation between the decoherence rates of all transitions.
(c) The couplings can also be varied along the noise insensitive diagonal feature of the coupler spectrum, which is optimal for quantum simulation.
}
\label{fig:tunable_coupling}
\end{figure}

The interactions between the qubits can be tuned by changing the flux bias point of the coupler circuit.
The flux tuning landscape of the coupler is shown in \figref{fig:tunable_coupling}(a).
The dark, diagonal feature in the transmission spectrum indicates the flux manifold along which the coupler frequency gap is minimal \cite{menke2022inprep}.
It arises from the coupler excited state pushing the coupler resonator down in frequency via dispersive interaction.
Simulations of the circuit predict maximum 3-local coupling in this regime.

We identify two distinct tuning paths for the 3-local coupling between the qubits.
The first is a vertical ``on-off" path along $\Phi_{\text{C}2}$ with fixed $\Phi_{\text{C}1}$, which enables the maximum range of $K_{123}$ including zero coupling.
The 3-local coupling is tuned from 0.8\,MHz to -4.6\,MHz along this path, as seen in the visualization of the extracted couplings in \figref{fig:tunable_coupling}(b).
These Hamiltonian parameters are extracted using the estimation procedure that was described above at each flux point.
We also observe that the local qubit fields and 2-local interactions between the qubits are modified by the coupler.
When using the coupler for practical applications, additional 2-local couplers for each qubit pair can be used to eliminate such spurious couplings and tune all Hamiltonian parameters independently \cite{weber2017coherent}.
The qubit decoherence rates are largest in the steepest region of the coupler spectrum.
Higher coherence is recovered at the point of maximum 3-local coupling, which is a flux insensitive point of the coupler.

The interactions in the 3-qubit system are a combination of two contributions:
first, the computational states of the isolated qubit circuits interact directly, which leads to 2-local interactions in the effective spin model.
Second, capacitive and inductive interactions between the computational and higher excited modes of the circuit modify both the 2-local and 3-local interactions.
As a result, a small 3-local coupling of 0.51\,MHz is present even when the coupler is turned off and the coupler excited state is far detuned ($\Phi_{\text{C}2} = 0\,\Phi_0$).
When the coupler is turned on ($\Phi_{\text{C}2} \sim 0.5\,\Phi_0$), its frequency gap drops to about 9.5\,GHz and the coupler excited state interacts with nearby qubit modes.
Numerical simulations of the full circuit Hamiltonian reveal an approximate picture of the higher-excited state frequencies, which is detailed in the Supplementary Information.
Repulsion or attraction between the coupler excited state and the qubit modes, most prominently the computational states $\ket{101}$ and $\ket{110}$, modify the effective spin Hamiltonian of the system.
As a result, the coupling strengths in the effective spin system are modified.
The tuning rate is highest when the coupler flux is close to $0.5\,\Phi_0$, which is when the coupler dispersion is steepest and its frequency is lowest.
The measured coupling parameters in \figref{fig:tunable_coupling}(b) reflect this tuning behavior, which validates our understanding of the multi-mode system.

The second tuning path follows the diagonal feature, and it enables tunability of $K_{123}$ between -3.2\,MHz and -6.5\,MHz.
It preserves maximum coherence of the qubits as the coupler stays first-order insensitive to flux noise at these fluxes.
The stable coherence is evident in the bottom panel of \figref{fig:tunable_coupling}(c).
In addition, all the interactions tune smoothly, and the change in spurious 2-local coupling is smaller than in the on-off tuning direction.
As a result, the coherent tuning regime is suitable for variations of the coupling during analog simulations or for digital gates, whereas the on-off tuning path is optimal for quickly turning the coupling on or off during an annealing protocol.

In conclusion, we have demonstrated a coherent quantum system that exhibits tunable multi-body interactions.
Multi-qubit effects are of fundamental interest as they do not arise naturally in non-relativistic quantum systems, and they provide a resource for analog quantum simulation, problem Hamiltonian engineering for quantum annealing, and gate design for digital algorithms.
In our demonstration, the system Hamiltonian is estimated via digital multi-qubit pulse sequences that precisely determine the frequencies of all computational eigenstates, enabling the distinct identification of 3-local interactions in the presence of lower-order couplings.
We study two different tuning regimes for the 3-local coupling and identify interactions between higher-excited states of the circuit as the source of the effective interaction.
The coherence times of the qubits, which we expect to be limited by flux noise from on-chip sources \cite{braumuller2020characterizing}, can be improved by reducing the loop size of the flux qubits and coupler, coupling the circuit elements galvanically rather than by a mutual inductance.
We highlight that the coupling scheme is compatible with other qubit modalities such as the transmon \cite{koch2007charge} or fluxonium qubit \cite{manucharyan2009fluxonium} and thus could serve as an efficient resource for gate-model quantum applications.
The studied superconducting circuit includes an additional flux qubit, which can be used to demonstrate 4-body interactions.
Moreover, the coupling scheme is extensible to even higher orders of interactions by adding additional loops to the coupler circuit \cite{menke2022inprep}.

It is a pleasure to thank X. Dai and F. Wilhelm-Mauch for valuable discussions, D. K. Kim for assistance with the fabrication, and J. I-J. Wang for taking the optical micrograph of the device.
This research was funded in part by the Office of the Director of National Intelligence (ODNI), Intelligence Advanced Research Projects Activity (IARPA) under Air Force Contract No. FA8702-15-D-0001. The views and conclusions contained herein are those of the authors and should not be interpreted as necessarily representing the official policies or endorsements, either expressed or implied, of the ODNI, IARPA, or the U.S. Government.


\bibliographystyle{apsrev4-1}
\bibliography{main}

\end{document}


\title{SUPPLEMENTARY INFORMATION\\ Demonstration of tunable three-body interactions between superconducting qubits}

\author{Tim Menke}
\email{timmenke@mit.edu}
\thanks{These authors contributed equally to this work.}
\affiliation{\rle}
\affiliation{\mitphysics}
\affiliation{\harvard}

\author{William P. Banner}
\thanks{These authors contributed equally to this work.}
\affiliation{\miteecs}

\author{Thomas R. Bergamaschi}

\author{Agustin Di Paolo}
\affiliation{\rle}

\author{Antti Veps\"al\"ainen}
\affiliation{\rle}

\author{Steven J. Weber}
\affiliation{\mitll}

\author{Roni Winik}
\affiliation{\rle}

\author{Alexander Melville}
\affiliation{\mitll}

\author{Bethany M. Niedzielski}
\affiliation{\mitll}

\author{Danna Rosenberg}
\affiliation{\mitll}

\author{Kyle Serniak}
\affiliation{\mitll}

\author{Mollie E. Schwartz}
\affiliation{\mitll}

\author{Jonilyn L. Yoder}
\affiliation{\mitll}

\author{Al\'an Aspuru-Guzik}
\affiliation{\ut}
\affiliation{\vectorinst}
\affiliation{\cifar}

\author{Simon Gustavsson}
\affiliation{\rle}

\author{Jeffrey A. Grover}
\affiliation{\rle}

\author{Cyrus F. Hirjibehedin}
\affiliation{\mitll}

\author{Andrew J. Kerman}
\affiliation{\mitll}

\author{William D. Oliver}
\email{william.oliver@mit.edu}
\affiliation{\mitphysics}
\affiliation{\rle}
\affiliation{\miteecs}
\affiliation{\mitll}

\date{\today}

\maketitle

\section{Experimental setup}

\begin{table}[h!]
    \begin{tabular}{ccc}
    \hlinewd{1.5pt}
        Component             & Manufacturer       & Model   \\ \hline
        Dilution Refrigerator\hspace{1em} & Leiden Cryogenics\hspace{1em}  & CF-CS81-1500-Maglev \\
        Network Analyzer      & Agilent            & N5232A PNA-L \\
        RF Sources            & Rohde \& Schwarz   & SGS100  \\
        DC Sources            & QDevil             & QDAC    \\
        Control Chassis       & Keysight           & M9019A  \\
        AWGs                  & Keysight           & M3202A  \\
        ADC                   & Keysight           & M3102A  \\
    \hlinewd{1.5pt}
    \end{tabular}
\caption{\textbf{Summary of control equipment.} The manufacturers and model numbers of the control equipment used for the experiment.}
\label{tab:equipment}
\end{table}

\begin{figure}[ht]
    \centering
    \includegraphics[scale=1.0]{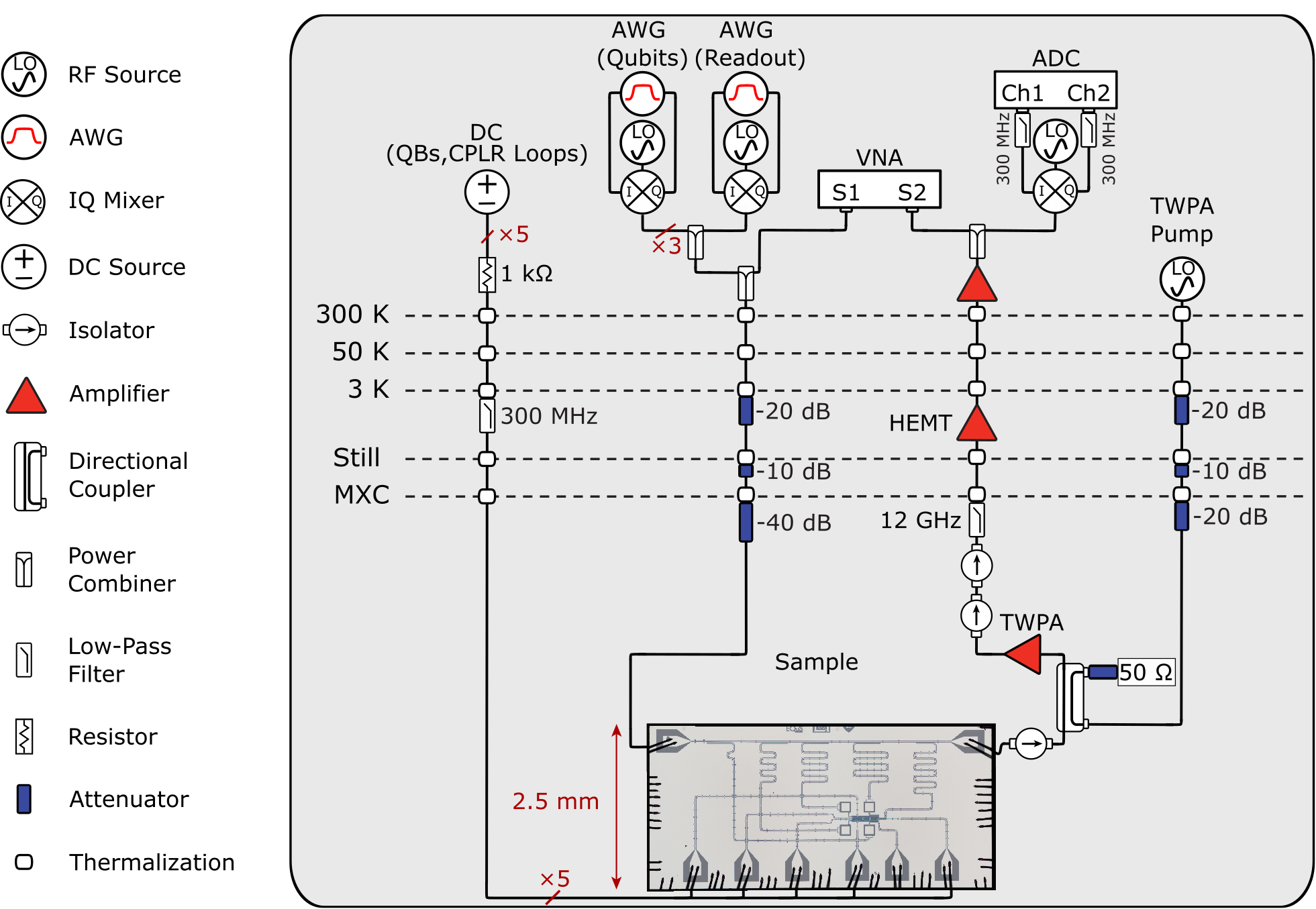}
    \caption{
        Wiring diagram of the experiment, including control electronics at room temperature as well as the setup inside the dilution refrigerator.
    }
    \label{fig:wiring_diagram}
\end{figure}

The experiment is performed in a Leiden Cryogenics CF-CS81-1500-Maglev dilution refrigerator, which is operated at 8\,mK as measured at the mixing chamber stage.
The setup of the experiment is visualized in \figref{fig:wiring_diagram}, and the manufacturers and models of the equipment used throughout the experiment are summarized in Table~\ref{tab:equipment}.
Four separate Keysight PXI arbitrary waveform generators (AWGs) shape the microwave pulses for control (3 AWGs) and readout (1 AWG).
The in-phase and quadrature components of the pulses are combined and upconverted using four individual Rhode \& Schwarz SGMA RF sources with built-in IQ mixers.
The qubit and readout signals are then combined using a series of combiners.
In parallel, an Agilent PNA-L vector network analyzer (VNA) is used in transmission-mode for crosstalk and readout calibration.
The VNA output is combined with the qubit and readout signals using a combiner.
After signal combination, the signals enter the fridge and are attenuated at each temperature stage to minimize thermal noise.
All measurements performed on-chip are transmission measurements through the feedline on the chip.

After the output port of the chip, the signal is isolated and amplified using a travelling wave parametric amplifier (TWPA), driven by a pump tone from an Agilent PSG signal generator at room temperature~[1,\,2].
Following TWPA amplification, the output signals are further isolated and filtered before additional amplification via a high electron mobility transistor (HEMT) amplifier and a room temperature Miteq amplifier.
At room temperature, a splitter diverts half of the output signal power to the VNA for calibration measurements.
The other half is down-converted using the local oscillator signal from the readout RF source and a Marki mixer.
Finally, the resulting in-phase and quadrature components of the signal are filtered, amplified, and digitized.
All qubit flux biasing is performed using on-chip local flux lines connected to filtered QDevil DC voltage sources with current-limiting 1\,k$\Omega$ resistors at the outputs.
    
\section{Hamiltonian description and simulation of the circuit}

\begin{figure}[ht]
    \centering
    \includegraphics[scale=0.7]{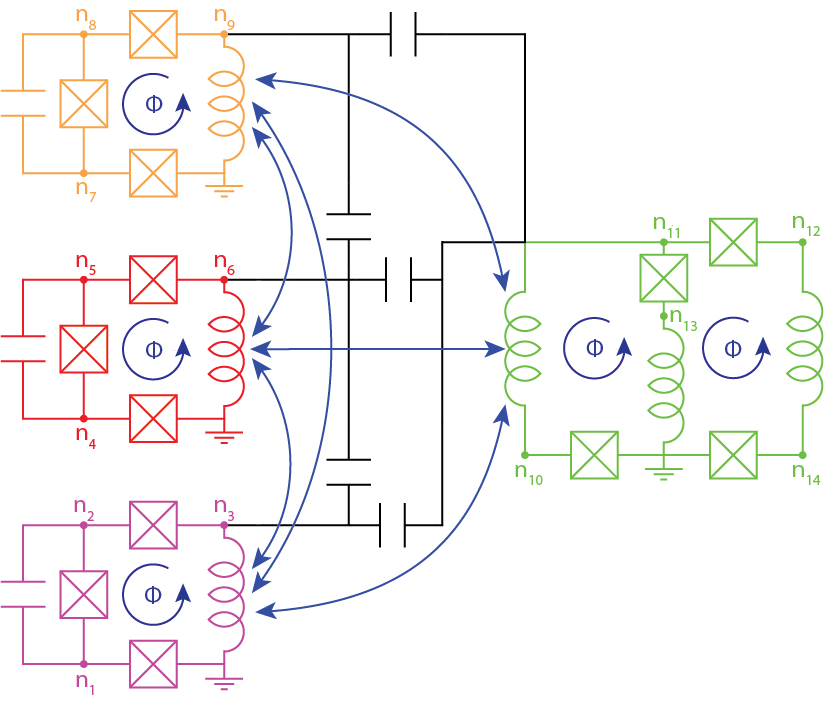}
    \caption{
        Simulated circuit diagram of the system. The circuit nodes are labelled as $n_1 ... n_{14}$. Both capacitive (black) and inductive (blue) couplings are simulated.
    }
    \label{fig:circuit_diagram}
\end{figure}

For quantum simulations of the circuit, we use the methods described in Ref.~[3].
In this way, we estimate both the properties of the isolated qubits and coupler as well as the eigenspectrum of the full circuit.
We include three qubits and the coupler in the simulation, ignoring the fourth qubit that is present on the chip but tuned far away in frequency.
The full circuit with all parameters and active nodes (labeled $n_1, ..., n_{14}$) is shown in \figref{fig:circuit_diagram}, which provides additional detail over Fig.~1(b) in the main text.
As the simulation of the 14-node circuit would be computationally infeasible otherwise, the individual qubits and coupler are diagonalized first and then coupled together in a hierarchical diagonalization procedure~[3].

\section{Qubit properties}

The properties of the qubits used in the experiment are summarized in Table~\ref{table:Ts}, with the measured qubit always biased at its sweetspot.
In the leftmost (individual) column, all qubits except the measured qubit are far detuned, as is the coupler.
In the center (uncoupled) column, all qubits are tuned to their flux-insensitive points and are therefore closer in frequency.
The coupler, however, remains far detuned.
In the rightmost (coupled) column, all qubits and the coupler are tuned to their operating points.

Looking at the trend from left to right in the table, we observe that the qubit $T_1$ times remain largely unaffected by the presence of the other qubits and the coupler.
Qubit $T_2$ times are approximately halved by when the other qubits are tuned to their flux insensitive points.
However, the presence of the coupler does not further limit this dephasing time to a significant degree.

\begin{table}
    \begin{tabular}{r c c c c c c c c c c c c}
        \hlinewd{1.5pt}
        \\[-0.9em]
        &&Individual&&&& Uncoupled&&&&  Coupled\\
        \\[-0.9em]
        && $T_1 (\text{ns})$ && $T_2 (\text{ns})$ && $T_1 (\text{ns})$ && $T_2 (\text{ns})$ && $T_1 (\text{ns})$&& $T_2 (\text{ns})$\\
        \\[-0.8em]
        \hline
        \\[-0.9em]
        Qubit 1 && $411 \pm 16$ && $288 \pm 20$ && $410 \pm 17$ && $135 \pm 17$ && $467 \pm 6$ && $142 \pm 28$\\
        \\[-0.9em]
        \hline
        \\[-0.9em]
        Qubit 2 && $450 \pm 16$ && $448 \pm 20$ && $442 \pm 17$ && $116 \pm 17$ && $338 \pm 6$ && $95 \pm 28$\\
        \\[-0.9em]
        \hline
        \\[-0.9em]
        Qubit 3 && $357 \pm 16$ && $334 \pm 20$ && $303 \pm 17$ && $235 \pm 17$ && $289 \pm 6$ && $146 \pm 28$\\
        \\[-0.9em]
        \hlinewd{1.5pt}
    \end{tabular}
    \caption{Longitudinal and transversal (Ramsey) decay rates at the sweetspot with all other qubits far detuned (individual), all other qubits at their respective sweetspots and coupler detuned (uncoupled), and all qubits and coupler at their respective sweetspots (coupled).}
    \label{table:Ts}
\end{table}
    
In order to determine the flux noise amplitude, we follow the procedure of Ref.~[4].
QB1 is tuned away from its flux-insensitive point and the pure dephasing rate ($\Gamma_\phi^E$) of the qubit is measured.
Under the assumption of Gaussian $1/f$ flux noise as the dominant noise source, the noise amplitude ($A_\phi$) can be extracted via the relation given in the following equation, where $\frac{\partial\omega}{\partial\Phi}$ is the slope of the qubit transition frequency ($\omega$) as a function of external flux ($\Phi$):

\begin{align}
    \Gamma^{E}_{\phi} = \sqrt{A_{\phi}\text{ln}2 } \left\lvert\frac{\partial\omega}{\partial\Phi}\right\lvert
    \label{eq:flux_noise}
\end{align}

In the study of Ref.~[4], the authors found that the flux noise amplitude depends on the wire thickness and perimeter of the SQUID.
The measured perimeters were not as large as the ones of our qubits, but we can extrapolate the findings.
As a result, we would expect a flux noise amplitude of about $5\,\mu\Phi_0$ for our qubits, which is about 5-6 times lower than the $27.2\,\mu\Phi_0$ that we measured.
We believe that the discrepancy results from the larger persistent current -- and therefore increased flux noise sensitivity -- and different geometry of the flux qubits used in this work.
The qubits may also experience additional flux noise due to their strong coupling to other qubits and the coupler.

\begin{figure}
    \centering
    \includegraphics[scale=1.0]{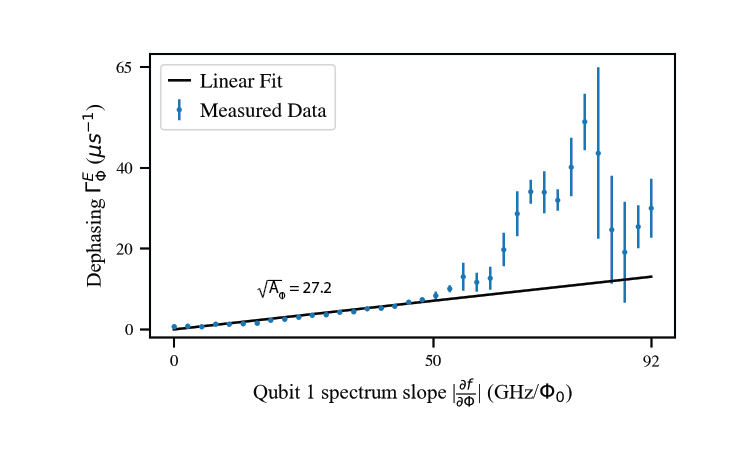}
    \caption{Dephasing rate as a function of the qubit flux dispersion. Under Gaussian noise assumptions, the extracted slope corresponds to the amplitude of the noise present in the qubit environment near the sweet spot.}
    \label{fig:flux_noise}
\end{figure}

\section{Coupler properties}

\begin{figure}[ht]
    \centering
    \includegraphics[scale=1.0]{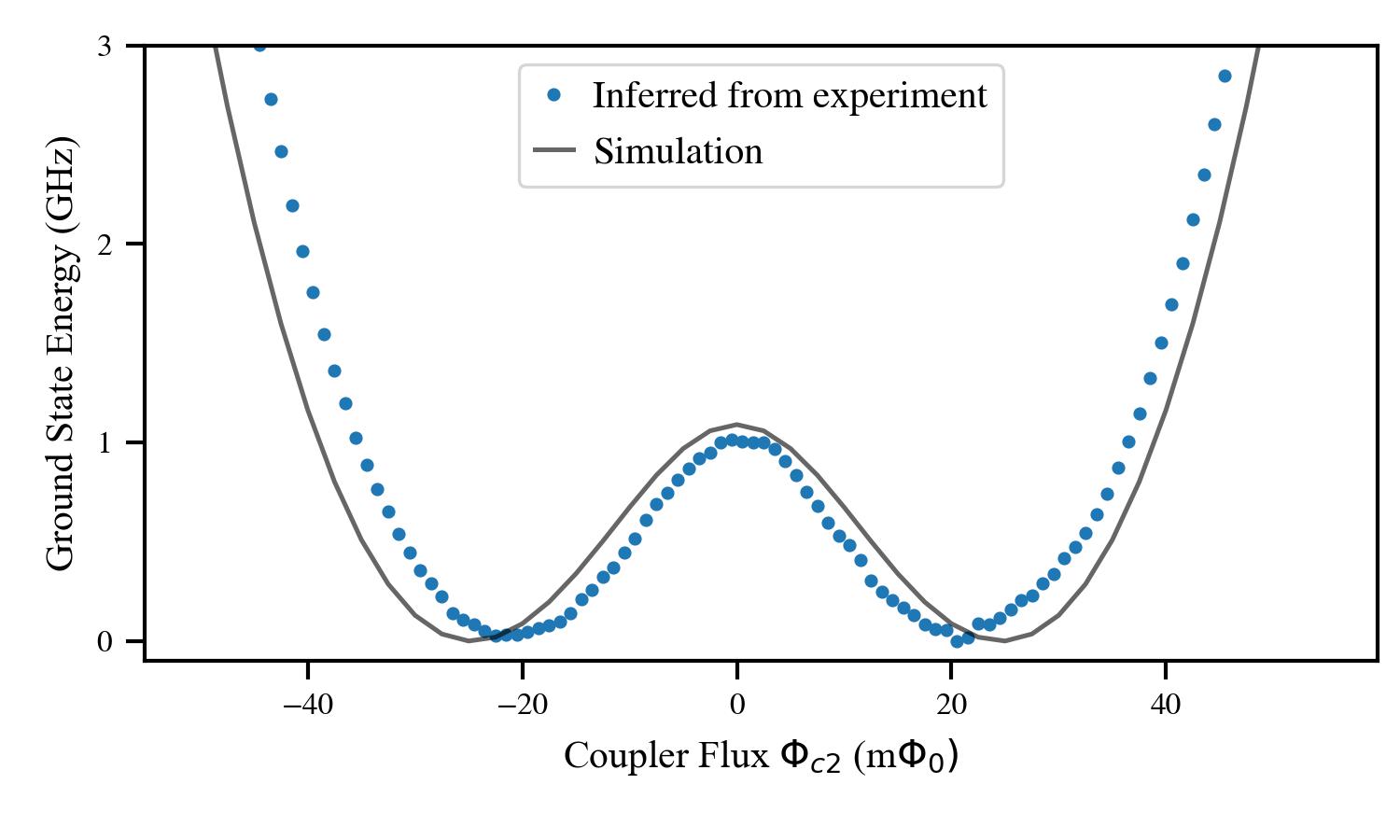}
    \caption{
        Ground state dispersion of the coupler versus flux. The designed, numerically simulated dispersion is shown in black, and the blue dots show the experimentally extracted dispersion up to an overall tilt.
    }
    \label{fig:coupler_ground}
\end{figure}

An analysis of the coupler spectrum is performed following the procedure outlined in Ref.~[5].
In this procedure, only a single qubit is tuned to its flux insensitive point, while all others remain far detuned.
The coupler flux is stepped such that the coupler crosses its the minimal gap.
At each step, the induced current in the qubit loop is calculated based on the shift in the qubit flux from the flux insensitive point.
We determine the slope of the coupler ground state from this shift.
Integrating the slope yields the ground state dispersion of the coupler up to a linear tilt, which is shown in \figref{fig:coupler_ground}.
The coupler dispersion has a double-well profile with a well-to-well spacing of approximately 40\,m$\Phi_0$ in $\Phi_{C1}$.
It matches well with the designed dispersion (green solid line in \figref{fig:coupler_ground}), which was engineered to generate multi-spin interactions.

\section{Flux crosstalk calibration}
    
We follow the procedure outlined in [6], which treats the flux crosstalk calibration as an iterative optimization problem.
The system crosstalk is represented as a set of linear equations that map voltage source settings to applied fluxes, assuming a linear voltage-current relation.
The inverse of the respective crosstalk matrix represents the correction that needs to be applied.
The relationship between fluxes and voltages is explicitly defined in equation \ref{eq:linear-xtalk-eq}, where $\mathbf{f}$ is the vector of fluxes in each loop, C is the crosstalk matrix, $\mathbf{V}$ is the vector of applied voltages, and $\mathbf{f}_0$ is the vector of flux offsets when no current is applied, which arises from spurious fields:
    
\begin{align}
    \mathbf{f} &= \text{C}\,\mathbf{V} + \mathbf{f}_0.
    \label{eq:linear-xtalk-eq}
\end{align}
    
\begin{figure}
    \centering
    \includegraphics[scale = 1.0]{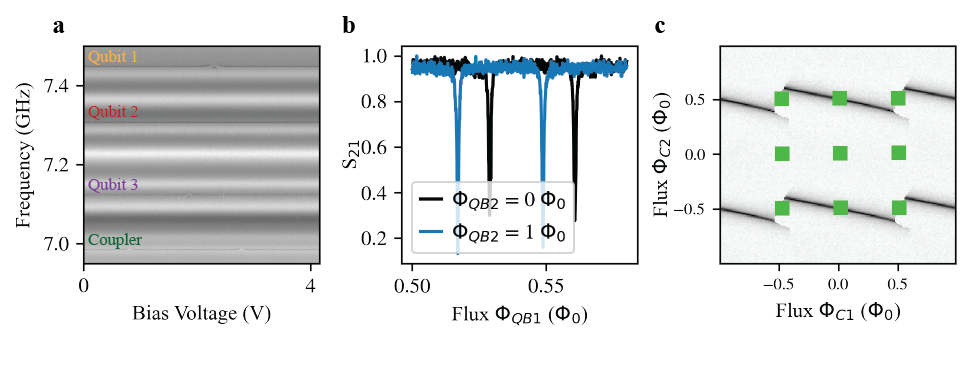}
    \caption{Flux crosstalk calibration. (a) A measurement across all four resonators to obtain an initial estimate of the bias voltage required to apply one flux quantum to each loop. The bias voltage is applied to a single flux antenna. (b) Transmission measurement versus $\Phi_{\text{QB}1}$ at a frequency just below the bare QB1 resonator frequency. The flux insensitive point is located between the two dips. The crosstalk from the QB2 flux is determined by stepping $\Phi_{\text{QB}1}$ by one flux quantum and determining the shift of the dips. (c) Coupler resonator transmission versus the two coupler fluxes. Image processing has been used to extract symmetry points (green). Crosstalk between the coupler fluxes is determined from the shear of the grid of symmetry points. In this example, the crosstalk is already well-calibrated.}
    \label{fig:xtalk}
\end{figure}

The crosstalk matrix C is determined using the natural periodicity of the flux-tunable elements and the shift that they impart on their respective resonator when they are close to it in frequency.
The periodicity, which is measured in units of the flux quantum $\Phi_0$, is first estimated for the qubits via resonator spectroscopy of all flux loops with respect to all antennas.
An example scan is shown in \figref{fig:xtalk}(a), where dips in the resonator frequency indicate the location of the periodically spaced flux insensitive points.
For the coupler, which consists of two strongly coupled loops, the periodicity needs to be determined from a 2d flux sweep:
The coupler resonator transmission is measured just below the bare resonator frequency while the coupler fluxes $\Phi_{\text{C}1}$ and $\Phi_{\text{C}2}$ are swept.
We thus obtain 2d data as shown in \figref{fig:xtalk}(c), which has also been shown in Fig.~4(a) in the main text.
At the beginning of the calibration, the arrangement of the diagonal features is sheared, and the crosstalk between the coupler fluxes is determined from the unit vectors between the diagonals.
After this initial estimate, a prescribed set of measurements follows:
    
\begin{enumerate}
    \item The resonator of each qubit is probed just below its bare frequency while the ``primary" flux of the qubit is swept. We label the flux from the antenna that is closest to the qubit as primary. An example for the data is shown in \figref{fig:xtalk}(b) (black trace), with the flux insensitive point located in the middle between the two dips. By finding two or more flux insensitive points, the periodicity of the qubit is revealed. The diagonal elements of C are determined from the qubit periodicities.
    
    \item The measurement from step (1) is repeated around a flux insensitive point while each non-primary flux in the flux vector is stepped by one flux quantum.
    We choose to step the non-primary fluxes by a full flux quantum in order to avoid second-order flux crosstalk from persistent currents of the qubits and coupler.
    The shift of the resonator scan (see \figref{fig:xtalk}(b)) determines the off-diagonal elements of C.
    
    \item The coupler resonator is measured at a single frequency in two dimensions across both $\Phi_{\text{C}1}$ and $\Phi_{\text{C}2}$ (see \figref{fig:xtalk}(c)).
    Using the image-processing method outlined in [6], the symmetry points of the measurement are extracted.
    The periodicity of the symmetry points and their vertical and horizontal alignment are used to extract the elements in the coupler block of the matrix C.
    
    \item The edge of the diagonal feature in the 2d coupler scan from step (3) is used as a calibration feature.
    Care needs be taken at this point as the feature is hysteretic.
    Stepping the qubit fluxes and determining the shift of the calibration feature reveals the final off-diagonal elements.
    
    \item The populated crosstalk matrix C is inverted to correct for the crosstalk present in the system.
    In particular, it is applied in addition to partial corrections that were determined previously.
    
    \item Steps (1-5) are repeated for several iterations, which successively improves the accuracy and removes second-order crosstalk effects.
\end{enumerate}
    
In this experiment, we performed six iterations of the described crosstalk calibration.
The dilution refrigerator was cycled between iterations four and five, which made the previous iterations less accurate and prompted two iterations of recalibration.
Using this procedure, we achieved crosstalk calibration errors of less than $3.4\%$ over one flux quantum and a mean error of 0.5\% between all non-primary antenna-loop pairs.

\section{Details about the Hamiltonian estimation method}

\begin{figure}
    \centering
    \includegraphics{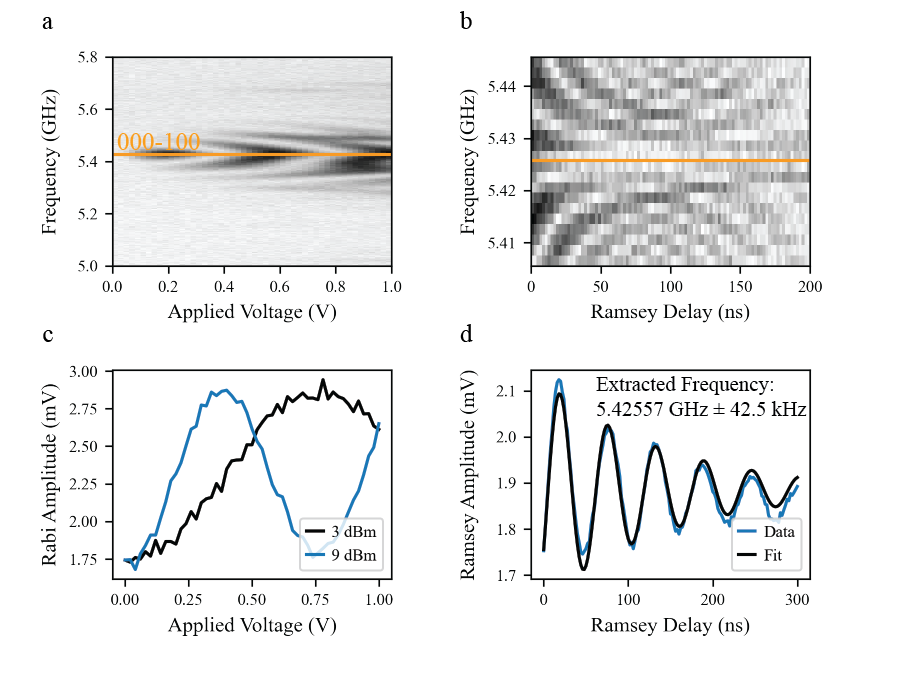}
    \caption{Characterization of transitions for the Hamiltonian estimation method. (a) Rabi measurement versus frequency, which is used to find candidate transitions. (b) Ramsey measurement versus frequency, which would reveal spurious transitions that interfere with the desired one. (c) Rabi measurement for two different drive powers. If the oscillation period is halved when the power is doubled as observed here, the candidate transition is confirmed to be a single-photon process. (d) Single Ramsey measurement detuned from the candidate transition, which is fit to find the precise frequency. All subplots represent the same $000-100$ transition, however readout drift in the system created different apparent Rabi and Ramsey amplitudes.}
    \label{fig:Transition_finding}
\end{figure}

The Hamiltonian estimation method requires a precisely characterized set of transition frequencies between the computational states.
A series of measurements is performed for each transition of interest in order to identify the transition and determine its frequency.
First, to obtain a set of candidate frequencies at which a transition of interest might exist, we perform a Rabi experiment versus drive frequency (see \figref{fig:Transition_finding}(a)).
For the lowest three transitions, this is a typical Rabi experiment.
For higher transitions, $\pi$-pulses must first be applied so that the proper transition can be accessed.
After candidate transitions are identified, Ramsey measurements are performed versus drive frequency in order to resolve any transitions which may be close in frequency (see \figref{fig:Transition_finding}(b)).
After confirming that transitions are not too close to be probed independently, Rabi experiments are performed on each candidate transition at multiple drive powers (see \figref{fig:Transition_finding}(c)).
For single-photon transitions, the Rabi oscillation period is halved for doubled power, whereas for two-photon transitions this period is quartered.
After confirming that a candidate transition is a single-photon transition, a highly averaged Ramsey measurement is performed at approximately $17\,$MHz detuning to precisely determine the transition frequency (see \figref{fig:Transition_finding}(d)).
The resulting Ramsey fringe oscillations are fit using \texttt{LMFIT}~[7], a nonlinear function fitting package, which provides precise error bounds based on confidence intervals.
All reported errors represent the fit error at one standard deviation from the mean.

For the Hamiltonian estimation method, only seven of the twelve possible transitions are required.
The seven transitions that are used in this experiment are the three bare qubit transitions ($000\rightarrow001$; $000\rightarrow010$; $000\rightarrow100$) as well as four higher-lying transitions ($001\rightarrow011$; $100\rightarrow101$; $100\rightarrow110$; $110\rightarrow111$). The frequencies of these transitions are related to the Hamiltonian parameters via a set of linear equations:
\begin{align*}
    \begin{bmatrix}
        f_{000-001}\\
        f_{000-010}\\
        f_{000-100}\\
        f_{001-011}\\
        f_{100-101}\\
        f_{100-110}\\
        f_{110-111}
    \end{bmatrix}
    & =
    \begin{bmatrix}
        \phantom{-}0 & \phantom{-}0 & \phantom{-}1 & \phantom{-}0 & -2 & -2 & -2\phantom{-} \\
        \phantom{-}0 & \phantom{-}1 & \phantom{-}0 & -2 & \phantom{-}0 & -2 & -2\phantom{-} \\
        \phantom{-}1 & \phantom{-}0 & \phantom{-}0 & -2 & -2 & \phantom{-}0 & -2\phantom{-}\\
        \phantom{-}0 & \phantom{-}1 & \phantom{-}0 & -2 & \phantom{-}0 & \phantom{-}2 & \phantom{-}2\phantom{-} \\
        \phantom{-}0 & \phantom{-}0 & \phantom{-}1 & \phantom{-}0 & \phantom{-}2 & -2 & \phantom{-}2\phantom{-} \\
        \phantom{-}0 & \phantom{-}1 & \phantom{-}0 & \phantom{-}2 & -2 & \phantom{-}2 & \phantom{-}0\phantom{-} \\
        \phantom{-}0 & \phantom{-}0 & \phantom{-}1 & \phantom{-}0 & \phantom{-}2 & \phantom{-}2 & -2\phantom{-}
    \end{bmatrix}
    \begin{bmatrix}
        \omega_1\\
        \omega_2\\
        \omega_3\\
        J_{12}\\
        J_{13}\\
        J_{23}\\
        K_{123}
    \end{bmatrix}
    \label{eq:ham_params}
\end{align*}

As a final validation of the estimation method, an overcomplete set of all 12 transitions was measured. All possible complete sets of seven transitions, 384 in total, were used to extract the Hamiltonian parameters, which were then compared. As is shown in Table~\ref{table:Parameters}, this \emph{selection error}, the standard deviation of the extracted parameters for different sets of seven transitions, is on the same order as the estimated fit error. This is intuitive, as the main contribution to the error in the transition estimates is expected to be the fit error.

\begin{table}
    \begin{tabular}{lclclclclclclclcl}
        \hlinewd{1.5pt}
        \\[-0.9em]
        Transitions && $f_{000-001}$ && $f_{000-010}$ && $f_{000-100}$ && $f_{001-011}$ && $f_{100-101}$ && $f_{100-110}$ && $f_{110-111}$\\
        \\[-0.9em]
        && $2.58774$ GHz 
        && $4.62194$ GHz 
        && $5.42518$ GHz 
        && $5.18069$ GHz 
        && $2.59436$ GHz 
        && $4.57770$ GHz 
        && $3.18918$ GHz\\
        \\[-0.9em]
        && $\pm 40$ kHz
        && $\pm 40$ kHz 
        && $\pm 40$ kHz 
        && $\pm 20$ kHz 
        && $\pm 40$ kHz 
        && $\pm 20$ kHz 
        && $\pm 30$ kHz\\
        \\[-0.9em]
        \hline
        \\[-0.9em]
        Parameters && $\omega_1$ && $\omega_2$ && $\omega_3$ && $J_{12}$ && $J_{13}$ && $J_{23}$ && $K_{123}$\\
        \\[-0.9em]
        && $5.41538$ GHz 
        && $4.88821$ GHz 
        && $2.87944$ GHz 
        && $-6.55$ MHz 
        && $6.16$ MHz 
        && $144$ MHz 
        && $-4.51$ MHz\\
        \\[-0.9em]
        && $\pm 100$ kHz 
        && $\pm 40$ kHz 
        && $\pm 50$ kHz 
        && $\pm 20$ kHz 
        && $\pm 30$ kHz 
        && $\pm 20$ kHz 
        && $\pm 20$ kHz\\
        \\[-0.9em]
        \hline
        \\[-0.9em]
        Selection Error
        && $\pm 330$ kHz 
        && $\pm 310$ kHz 
        && $\pm 290$ kHz 
        && $\pm 110$ kHz 
        && $\pm 100$ kHz 
        && $\pm 100$ kHz 
        && $\pm 80$ kHz\\
        \\[-0.9em]
        \hlinewd{1.5pt}
    \end{tabular}
    \caption{An overview of the transition frequencies and extracted Hamiltonian parameters when the coupler is on. In addition, the selection error quantifies the deviation of the extracted values when different sets of transitions are used for the estimation.}
    \label{table:Parameters}
\end{table}

\section{Multi-level model of the coupling mechanism}
    
\begin{figure}
    \centering
    \includegraphics[scale=0.45]{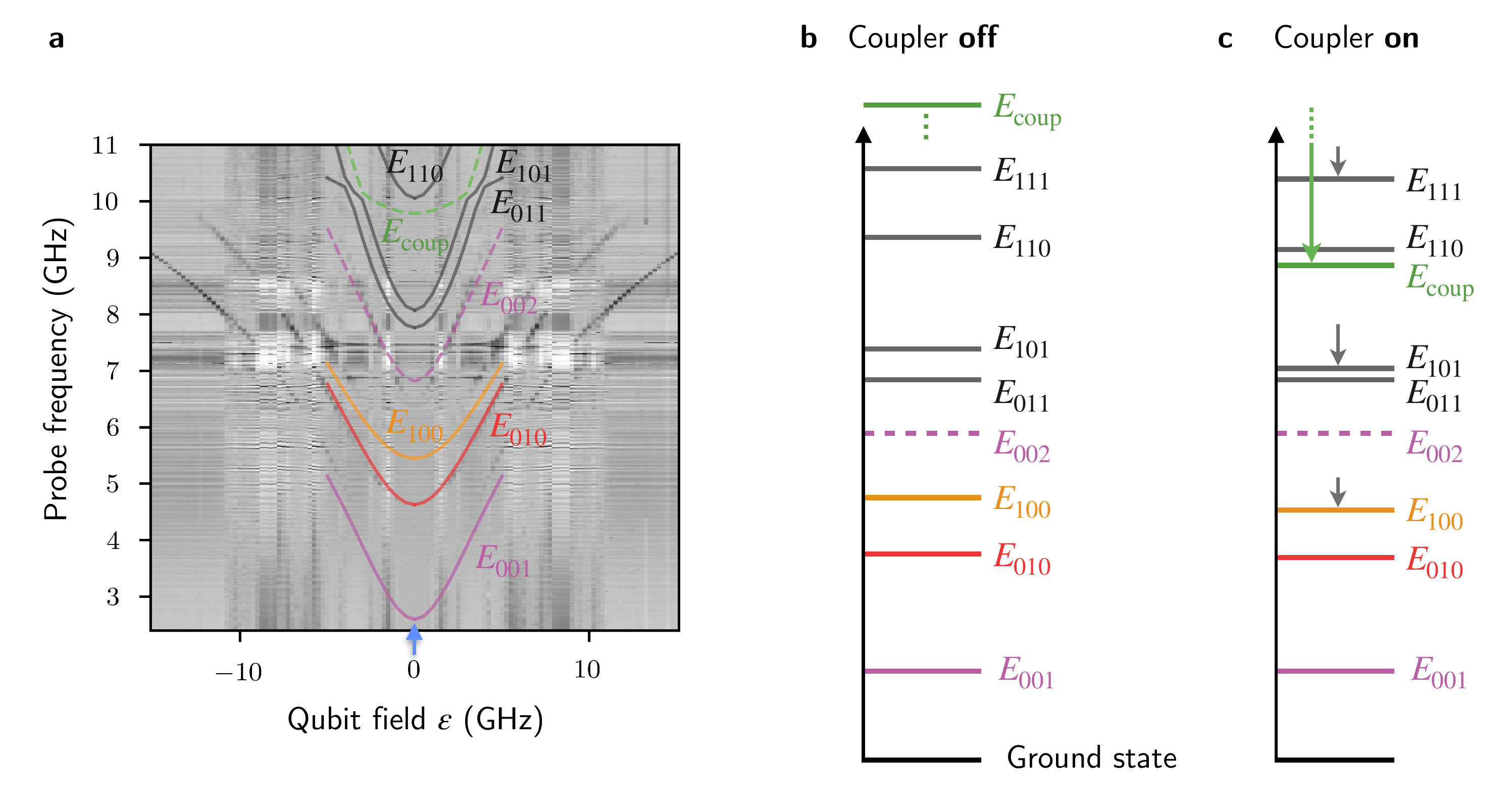}
    \caption{Multi-level model of the coupling mechanism. (a) Spectroscopy of the system with the coupler on. Eigenenergies from a full circuit Hamiltonian model are overlaid. The blue arrow indicates that the model is optimized at the flux insensitive point of the qubits. (b) Eigenenergies of the model with the coupler turned off. Higher excited states of the flux qubits, such as the $E_{002}$ state, induce a shift of select computational states and lead to effective 2- and 3-local interactions. (c) When the coupler is turned on, its excited state induces additional shifts of the computational states and thus tunes the 2- and 3-local interactions.}
    \label{fig:coupling_mechanism}
\end{figure}

The effective interactions in the 3-qubit system arise from two separate effects:
First, there is a direct interaction between the computational states of the isolated qubit circuits, which leads to 2-local interactions in the effective spin model.
Second, capacitive and inductive interactions between the computational and higher excited states of the circuit modify both the 2-local and 3-local interactions.
As described in the main text, a small 3-local coupling of 0.51\,MHz is present even when the coupler is turned off ($\Phi_{\text{C}1} = \Phi_{\text{C}2} = 0\,\Phi_0$).
At this bias point, the coupler ground-to-excited state energy gap is estimated in simulations to be tens of GHz, and the qubits are not affected by the coupler states.
Instead, the remaining 3-local coupling is the result of the second and third excited states of the flux qubits interacting with computational states that have multiple excitations.
Numerical simulations of the full circuit Hamiltonian were used to reconstruct an approximate picture of the mode frequencies beyond the directly measured states (see \figref{fig:coupling_mechanism}(b)).
    
When the coupler is turned on ($\Phi_{\text{C}2} \sim 0.5\,\Phi_0$), its gap drops to about 9\,GHz and the coupler excited state interacts with nearby qubit modes.
Some hybridization between the coupler and qubit modes occurs, but we have verified in Rabi and Ramsey experiments that the computational states retain their qubit characteristics.
In \figref{fig:coupling_mechanism}(c), it is shown how the eigenstates of the system are modified by the coupler.
The eigenenergies $E_{100}$, $E_{101}$ and $E_{111}$ are most affected.
As a result, the Hamiltonian parameters of the effective spin system are tuned, and we expect this effect to generate the larger 3-local coupling.
The circuit simulations here are not able to fully model all of the eigenenergies of the system, which is likely caused by simplifications such as Hilbert space truncation and removal of the fourth qubit from the circuit model.
A quantitatively accurate model would be obtained if $E_{101}$ was pushed down a few hundred MHz further by the coupler.
However, the circuit model and resulting energy level structure has provided important insights into the coupling mechanisms of the system.

\section*{Supplementary References}

\begin{itemize}

    \item[{[1]}] O. Yaakobi, L. Friedland, C. Macklin, and I. Siddiqi, ``Parametric amplification in Josephson junction embedded transmission lines," Phys. Rev. B \textbf{87}, 144301 (2013).
    
    \item[{[2]}] C. Macklin, K. O?Brien, D. Hover, M. E. Schwartz, V. Bolkhovsky, X. Zhang, W. D. Oliver, and I. Siddiqi, ``A near?quantum-limited Josephson traveling-wave parametric amplifier," Science \textbf{350}, 307 (2015).

    \item[{[3]}] A. J. Kerman, ``Efficient numerical simulation of complex Josephson quantum circuits," arXiv preprint arXiv:2010.14929 (2020).
    
    \item[{[4]}] J. Braum\"uller, L. Ding, A. P. Veps\"al\"ainen, Y. Sung, M. Kjaergaard, T. Menke, R. Winik, D. Kim, B. M. Niedzielski, A. Melville, \textit{et al.}, ``Characterizing and optimizing qubit coherence based on squid geometry," Physical Review Applied \textbf{13}, 054079 (2020).
    
    \item[{[5]}] T. Menke \textit{et al.}, ``Quantum landscape engineering of the superconducting circuit flux dispersion," in preparation.
    
    \item[{[6]}] X. Dai, D. M. Tennant, R. Trappen, A. J. Martinez, D. Melanson, M. A. Yurtalan, Y. Tang, S. Novikov, J. A. Grover, S. M. Disseler, J. I. Basham, R. Das, D. K. Kim, A. J. Melville, B. M. Niedzielski, S. J. Weber, J. L. Yoder, D. A. Lidar, and A. Lupascu, ``Calibration of flux crosstalk in large-scale flux-tunable superconducting quantum circuits," PRX Quantum \textbf{2}, 040313 (2021).
    
    \item[{[7]}] M. Newville, T. Stensitzki, D. B. Allen, M. Rawlik, A. Ingargiola, and A. Nelson, ``LMFIT: Non-linear least-square minimization and curve-fitting for python," Astrophysics Source Code Library , ascl (2016).
    
\end{itemize}
